\documentclass[twocolumn,floatfix,superscriptaddress,amsmath,showpacs,showkeys,aps,prb]{revtex4}

\usepackage[final]{graphicx}
\usepackage{t1enc}
\usepackage{times}

\begin{document}

\bibliographystyle{prsty}

\title{The Spin Reorientation Transition and Phase Diagram of Ultrathin
Ferromagnetic Films}

\author{Marianela Carubelli}
\email{carubell@famaf.unc.edu.ar}
\affiliation{Facultad de
Matem\'atica, Astronom\'{\i}a y F\'{\i}sica, Universidad Nacional
de C\'ordoba, \\ Ciudad Universitaria, 5000 C\'ordoba, Argentina}
\altaffiliation{Member of CONICET, Argentina}
\author{Orlando V. Billoni}
\email{billoni@famaf.unc.edu.ar} \affiliation{Facultad de
Matem\'atica, Astronom\'{\i}a y F\'{\i}sica, Universidad Nacional
de C\'ordoba, \\ Ciudad Universitaria, 5000 C\'ordoba, Argentina}
\altaffiliation{Member of CONICET, Argentina}
\author{Santiago Pigh\'{i}n}
\email{spighin@famaf.unc.edu.ar} \affiliation{Facultad de
Matem\'atica, Astronom\'{\i}a y F\'{\i}sica, Universidad Nacional
de C\'ordoba, \\ Ciudad Universitaria, 5000 C\'ordoba, Argentina}
\altaffiliation{Member of CONICET, Argentina}
\author{Sergio A. Cannas}
\email{cannas@famaf.unc.edu.ar} \affiliation{Facultad de
Matem\'atica, Astronom\'{\i}a y F\'{\i}sica, Universidad Nacional
de C\'ordoba, \\ Ciudad Universitaria, 5000 C\'ordoba, Argentina}
\altaffiliation{Member of CONICET, Argentina}
\author{Daniel A. Stariolo}
\email{stariolo@if.ufrgs.br}
\affiliation{Departamento de F\'{\i}sica,
Universidade Federal do Rio Grande do Sul\\
CP 15051, 91501--979, Porto Alegre, Brazil}
\altaffiliation{Research Associate of the Abdus Salam International Centre for Theoretical Physics,
Trieste, Italy}
\author{Francisco A. Tamarit}
\email{tamarit@famaf.unc.edu.ar} \affiliation{Facultad de  Matem\'atica, Astronom\'{\i}a
y F\'{\i}sica, Universidad Nacional de C\'ordoba, \\ Ciudad Universitaria, 5000
C\'ordoba, Argentina} \altaffiliation{Member of CONICET, Argentina}

\date{\today}

\begin{abstract}
We show results from Monte Carlo simulations of a two dimensional
Heisenberg model for ultrathin films with perpendicular anisotropy.
A complete phase diagram is obtained as a function of anisotropy and
temperature, spanning a wide range of behavior. We discuss our results in
relation with experimental findings in different ultrathin films.
We observe and
characterize a line of Spin Reorientation Transitions . This
transition from out of plane stripe order to in plane ferromagnetic order
presents a paramagnetic gap in between in a finite region in parameter
space, as reported in experiments. For
large anisotropies direct transitions from a low temperature stripe
phase to a paramagnetic or tetragonal phase with dominant
perpendicular magnetization is observed, also in agreement with experiments.

We also show the phase diagram for a system without exchange, i.e. with
pure dipolar and
anisotropy interactions. It shows a similar behavior to the ferromagnetic
case with antiferromagnetic instead of stripe phases at low temperatures. A
Spin Reorientation Transition is also found in this case.
\end{abstract}


\maketitle

\section{Introduction}
\label{introduction}
In recent years magnetic behavior of ultrathin films has become of
great technological
 importance due to the applications in magnetic storage
 devices. As the sizes become smaller and smaller a detailed microscopic
 characterization of magnetization processes on the nanometer scale is mandatory.
Magnetic order in ultrathin ferromagnetic films is very complex due to
the
competition between exchange and dipolar interactions on different
length scales, together with a strong influence of shape and
magnetocristalline
anisotropies of the sample. These in turn are very susceptible to the  growth
conditions of the films~\cite{Portmann2006}. In the last 20 years a
considerable amount of experimental results on different aspects of
magnetism in ultrathin films have appeared. Nevertheless, after a
careful analysis of the literature it is difficult to reach general
conclusions even in seemingly basic things as the kind of magnetic
order at low temperatures. In view of this complexity, theoretical
work on simplified models and computer simulations are essential for
rationalizing and guiding new experimental work. In early experiments
on Fe/Cu(100) films, Pappas {\it et al}.~\cite{PaKaHo1990} and Allenspach {\it et al}.~\cite{AlBi1992}
observed a spin reorientation transition (SRT) from a region with
perpendicular magnetization to one with in-plane magnetization. In the
first experiment Pappas {\it et al}. found a gap with a complete loss of magnetization  in
between the perpendicular and in-plane phases. Two hypothesis were put
forward for the origin of the gap: a dynamic origin due the
compensation of perpendicular and in-plane anisotropies in the region
around the SRT and a static one based on previous theoretical work by Yafet {\it et al}.~\cite{YaGy1988} who predicted a striped magnetic phase as the true
ground state of ultrathin films with perpendicular anisotropy.
In the second
experiment Allenspach {\it et al}. discarded the possibility of a completely
vanishing magnetization in the vicinity of the SRT, but instead
observed the emergence of stripe magnetic order with a temperature
dependent stripe width in general agreement with Yafet predictions. In
their measurements no gap was observed between the perpendicular and
in-plane phases.
We will show that in fact this kind of behavior, with a direct SRT from a striped to a ferromagnetic in-plane state, is
present in a particular anisotropy-temperature region in the
phase diagram of our model. Furthermore, the thickness dependence of
the SRT temperature observed by Pappas {\it et al}. can be qualitatively
reproduced by our results on a single monolayer, after noting that the
anisotropy behaves as the inverse of the film thickness, as discussed below.

More
recently Won {\it et al}.~\cite{WoWuCh2005} studied the SRT as a function of
temperature and thickness in Fe/Ni/Cu(001) films. They found an
exponential decrease of stripe width on approaching the SRT and the
possibility of a paramagnetic gap between the out of plane stripe
phase and the in-plane ferromagnetic phase. The existence of the gap
was interpreted by the authors in terms of a crossover between typical
dipolar and
anisotropy lengths. They defined a Curie temperature as a function of
the dipolar length and depending on it being higher or lower than the
SRT temperature, a paramagnetic gap may or may not manifest in the
system. Indeed, we will show that also this kind of behavior with a SRT and a gap is
present in a particular anisotropy-temperature region in the
phase diagram of our model.
Although we were not able to test quantitatively the
phenomenological arguments of Won {\it et al}. because of our too small
working stripe width, their conclusions are completely consistent with
the scenario that emerges from our simulations. In yet another set of
important experiments Vaterlaus {\it et al}.~\cite{VaStMaPiPoPe2000} found
evidence of a two step disordering process. The films show stripe
phases at low temperatures which loose orientational order and
eventually evolve into a ``tetragonal liquid phase'' with short range
stripe order showing 90$^o$ rotational symmetry. This phase further
evolves in a continuous way towards a final paramagnetic phase. In
these experiments the films present strong perpendicular anisotropy
and no SRT is observed; magnetization is always out of plane. We will
show that this is also observed in our simulations in the parameter
region corresponding to strong anisotropy. In this region of the phase
diagram a direct transition from a stripe phase to a paramagnetic (or
tetragonal) phase is observed. In the region of strong anisotropy the
thermodynamic phases can be studied in the Ising limit. Detailed
ground state calculations~\cite{MaWhRoDe1995} and numerical
simulations
~\cite{CaStTa2004,CaMiStTa2006,RaReTa2006} have been done in
recent years and a successful picture of this region of the phase
diagram has emerged. In an extended region of temperatures and
anisotropies MacIsaac {\it et al}.~\cite{MaDeWh1998} have presented a phase
diagram of an Heisenberg model with dipolar and exchanged
interactions. Their phase diagram (figure 3 of their letter) is
similar to our present results. Nevertheless both diagrams differ in
an important result: while they obtained a SRT from a low temperature perpendicular stripe phase to an in-plane ferromagnet   at higher temperatures (at variance with most experiments),  our results show the
inverse trend, {\it i.e.}, from an in-plane ferromagnet at low temperature to perpendicular a stripe or paramagnetic phase at high temperatures, consistent with experimental results. Our SRT line is
supported by experimental as well as several theoretical arguments as
will be explained below.

The nature of the different phase transitions is a delicate issue and
several controversial results are spread in the literature. In the
present work we did not pursue to set in a definitive answer, but
nevertheless we added  new results to the old ones. In the
high
anisotropy limit our results regarding the nature of the
stripe-tetragonal phase are again consistent with similar simulations
in Ising systems which point to predominantly first order
transitions.
This result is again at variance with the continuous transition
reported by MacIsaac {\it et al}. on the
same region~\cite{MaDeWh1998}. At
intermediate anisotropies the same transition line gradually changes its behavior
and the transition seems to become continuous or weakly first
order in the region where a gap is observed around the SRT.
This behavior is similar to
the phenomenology observed recently in field theoretical models for thin films
with Langevin dynamics~\cite{NiSt2007,BaSt2007}. Another relevant aspect concerns the possible existence of an intermediate nematic phase as
predicted theoretically by Abanov {\it et al}.~\cite{AbKaPoSa1995} and
recently on
more general grounds by Barci {\it et al}.~\cite{BaSt2007} and characterized
in Langevin simulations by Nicolao {\it et al}.~\cite{NiSt2007} and also in Monte
Carlo simulation of an Ising model by Cannas {\it et al}.~\cite{CaMiStTa2006}.
In the rest of
this work we will refer to stripe phases regardless of the existence
of true long range positional order or only orientational order.
Besides the stripes-tetragonal or paramagnetic transition line, we
obtained convincing evidence for a  first order nature of the SRT line and
the continuous nature of the in-plane ferromagnetic-paramagnetic
transition,
as will be shown below.

Finally we also show a complete phase diagram of the pure dipolar
system with perpendicular uniaxial anisotropy. The phase diagram in this
limit is similar to the one with exchange interaction, the main difference
being the small slope of the SRT line as compared to the ferromagnetic
case and the antiferromagnetic nature of the low temperatures phases.

\section{Model and Methods}
\label{model}

We have performed Monte Carlo (MC) simulations on the usual model Hamiltonian for ultrathin
films with exchange, dipolar and perpendicular anisotropy on a square lattice of side
$L=40$:
\begin{widetext}
\begin{equation}
{\cal H} = -\delta \sum_{<i,j>} \vec{S}_i \cdot \vec{S}_j +
\sum_{(i,j)} \left[ \frac{\vec{S}_i \cdot \vec{S}_j }{r_{ij}^3} - 3 \,
\frac{(\vec{S}_i \cdot \vec{r}_{ij}) \; (\vec{S}_j \cdot \vec{r}_{ij})}{r_{ij}^5} \right]
- \eta \sum_{i} (S_i^z)^2
\label{hamiltoniano}
\end{equation}
\end{widetext}
where the exchange and anisotropy constants are normalized relative to the dipolar coupling
constant, $<i,j>$ stands for a sum over nearest neighbors pairs of sites in the lattice, $(i,j)$ stands for a sum over {\it all distinct} pairs and $r_{ij}\equiv |\vec{r}_i - \vec{r}_j|$ is the distance between spins $i$ and $j$.
All the simulations were done using the Metropolis algorithm and periodic boundary conditions
were imposed on the lattice by means of the Ewald sums technique. All the results
presented in section \ref{monolayer} refer to the case $\delta=3$ which corresponds, for large values of $\eta$, to a ground
state with out of plane stripe magnetic structure of width~\cite{MaWhRoDe1995} $h=4$ and to
an in-plane ferromagnetic ground state for small anisotropy (see figure \ref{fig1}).

Each spin is defined by a unit vector with components $S^x,S^y,S^z$. The phase diagram has
been obtained measuring the out plane magnetization:
\begin{equation}
 M_z \equiv \frac{1}{N}\sum_{\vec{r}} \left< S^z(\vec{r}) \right>,
\label{mz}
\end{equation}
the in-plane magnetization:
\begin{equation}
 M^{||} \equiv \sqrt{(M^x)^2 + (M^y)^2} ,
\end{equation}
and an orientational order parameter similar to the one defined by Booth {\it et al}.
~\cite{BoMaWhDe1995}:
\begin{equation}
 O_{hv} \equiv \left< \left| \frac{n_h-n_v}{n_h+n_v} \right| \right>
\end{equation}
\noindent where $\left< \cdots \right>$ stands for a thermal average, $n_h$ ($n_v$) is the number of horizontal (vertical)  pairs of nearest neighbor spins with antialigned perpendicular component, {\it i.e.},
\begin{equation}
 n_h = \frac{1}{2}\sum_{\vec{r}} \,\left\{1-sig\left[S^z(r_x,r_y),\,S^z(r_x+1,r_y)\right]\right\}
\end{equation}
\noindent and a similar definition for $n_v$, where $sig(x,y)$ is the sign of the product of $x$ and
$y$. In the previous definitions $N=L\times L$ is the number of spins and $M^x$, $M^y$ are
defined similarly to equation (\ref{mz}). Other quantities calculated were the specific heat

\begin{equation}\label{Cc}
    C \equiv \frac{1}{NT^2} \left( \left< H^2 \right>
-\left< H \right>^2\right)
\end{equation}

\noindent and the mean absolute magnetization

\begin{equation}\label{P}
    P \equiv \frac{1}{N} \sum_{\vec{r}} \left< \left|S^z(\vec{r}) \right|\right>
\end{equation}

In section \ref{delta0} we calculate the phase diagram for $\delta=0$ (dipolar  interactions plus anisotropy). In this case the relevant phases at low temperatures are antiferromagnetic (AF) . For high values of the anisotropy the ground state is AF with sublattice magnetization and all the spins oriented perpendicular to the plane\cite{MaWhDePo1996}. For low values of the anisotropy the ground state is a highly degenerated planar AF state; the different configurations of this state are described in Ref.\cite{DeMaBoWh1997}. To characterize the perpendicular AF state we calculated the staggered perpendicular magnetization

\begin{equation}\label{mstagger}
    M_{s\perp} \equiv \frac{1}{N} \left< \left| \sum_{\vec{r}} (-1)^{r_x+r_y} S^z(\vec{r})\right| \right>
\end{equation}

\noindent  To characterize the planar AF state we calculated the following orientational order parameter\cite{MaWhDePo1996,DeMaBoWh1997}

\begin{equation}\label{mstaggerp}
    M_{s\parallel} \equiv \frac{1}{N} \left< \left| \sum_{\vec{r}} (-1)^{r_y} S^x(\vec{r}) \, \hat{x} + (-1)^{r_x} S^y(\vec{r}) \, \hat{y} \right| \right>
\end{equation}

To obtain the phase diagrams $T$ vs $\eta$ we analyzed the behavior of the above quantities by fixing $\eta$ and varying $T$ or viceversa. Those curves were calculated using two different simulation protocols.

To analyze equilibrium properties we use a ladder protocol, where the system is initialized at some configuration close to the equilibrium one (either the corresponding ground state at low temperatures or the paramagnetic one at high temperatures) and the independent parameter ($\eta$ or $T$) is varied at discrete steps. The initial configuration for every value of the independent parameter was taken as the last one of the previous value; then  we discarded the first $t_e$ Monte Carlo Steps (MCS) needed for equilibration and calculated the averages over the next $t_m$ MCS. A MCS is
defined as a complete cycle of $N$ spin update trials, according
to the Metropolis  algorithm. Typical values of $t_e$ were around $10^5$ MCS, while typical values
of $t_m$ were between $10^3$ and $10^4$ MCS.

To analyze the possible existence of hysteresis effects we used a ``cooling-heating'' procedure, varying the temperature (or $\eta$)  according to a linear protocol $T(t)= T(0) \pm r\, t$, where $T(0)$ is the initial temperature, $t$ is measured in MCS and $r$ is a constant rate. Before starting the protocol we let the system to equilibrate during $t_e$ MCS from some appropriated initial configuration (as in the previous protocol) and then we recorded the quantities of interest as a function of time along a complete path to the final temperature; then we repeated the procedure several times, averaging the whole curves over different sets of initial configurations (in the cases were they are random) and over different sequences of thermal noise; typical sample sizes were between 50 and 100.

\section{Phase Diagram for the ferromagnetic monolayer}
\label{monolayer}

\begin{figure}
\begin{center}
\includegraphics[scale=0.35]{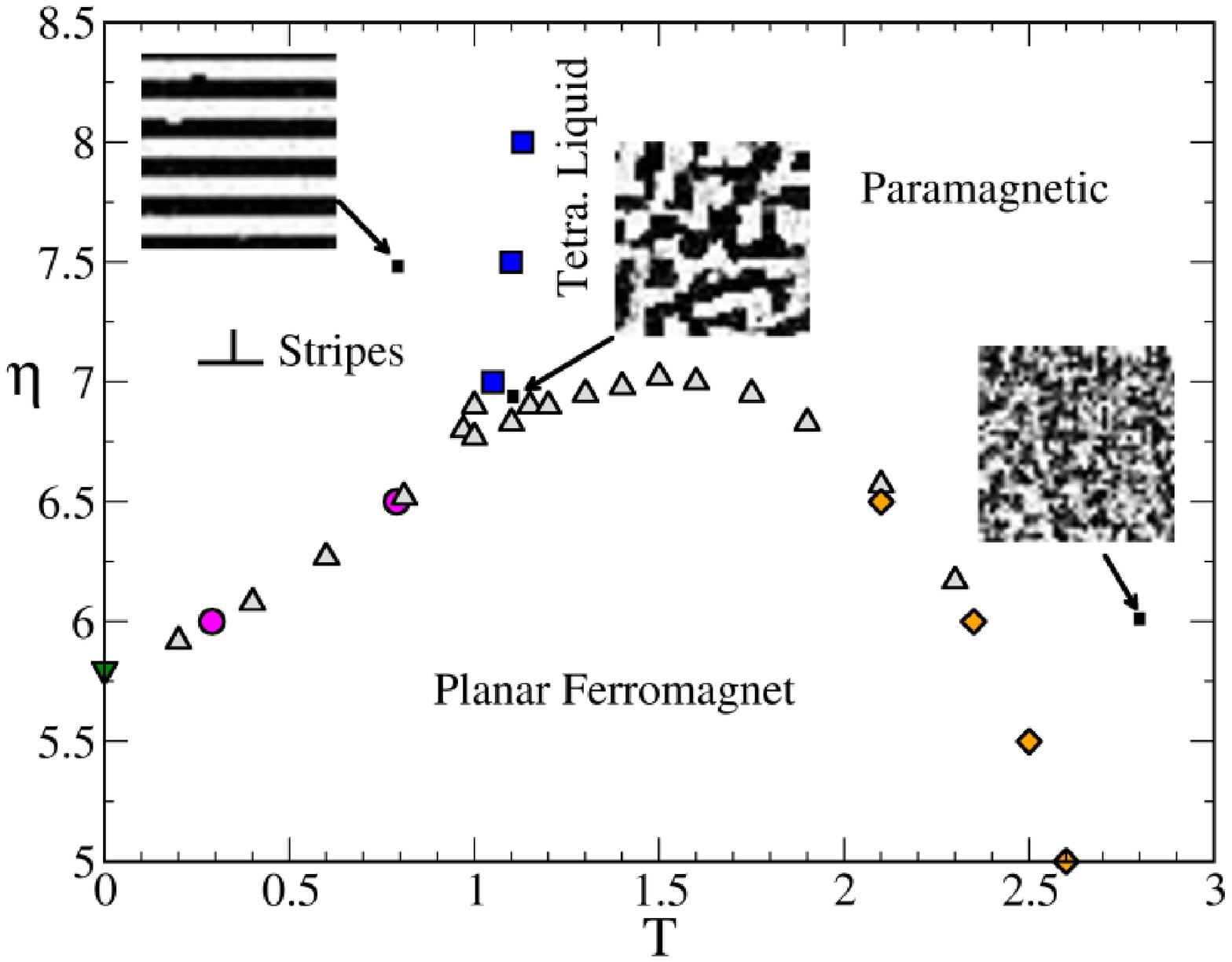}
\caption{(Color online) The phase diagram $\eta$ vs. Temperature for $\delta=3$.
The different symbols corresponds to different calculation methods; triangle down (green): ground state calculation; circle (red): energy histograms simulations; square (blue): order parameter histogram simulations; triangle up (white): equilibrium and non-equilibrium order parameters simulations (see methods); diamond (yellow): specific heat simulations. Some typical spin configurations (perpendicular component of the spins) are shown at different phases}
\label{fig1}
\end{center}
\end{figure}

In figure \ref{fig1} we show the phase diagram in the $(\eta,T)$ plane. The different
transition lines were obtained by measuring more than one quantity as is indicated in
the figure with different symbols.

As can be seen all lines are smooth which gives confidence to the quality of the data.
We can clearly distinguish three different phases:
perpendicular stripes for low temperatures and strong anisotropy, in-plane ferromagnet
for low temperatures and weak anisotropy and paramagnetic behavior for high temperatures. The disordering of
the stripe phase with temperature evolves through a region where the orientational order
is lost but a lower symmetry to $90^o$ rotations survives and continuously evolves to
the complete paramagnetic state. We will call this region {\em tetragonal phase} although
there is no clear evidence that a sharp transition to a paramagnetic phase with full rotational symmetry is
present at high temperatures.

\subsection{Stripes-Tetragonal transition}

\begin{figure}
\hspace{-1cm}\includegraphics[scale=0.3,angle=-90]{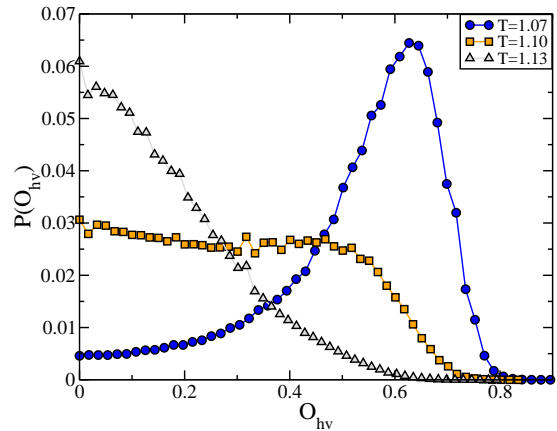}
\caption{(Color online) Order parameter $O_{h\nu}$ per spin histograms for $\eta=7.5$ and
different temperatures. The histograms were calculated for $30 \times 10^6$ values of the energy, measured along a single MC run.}
\label{fig2}
\end{figure}

\begin{figure}
\begin{center}
\hspace{-1cm}\includegraphics[scale=0.32,angle=-90]{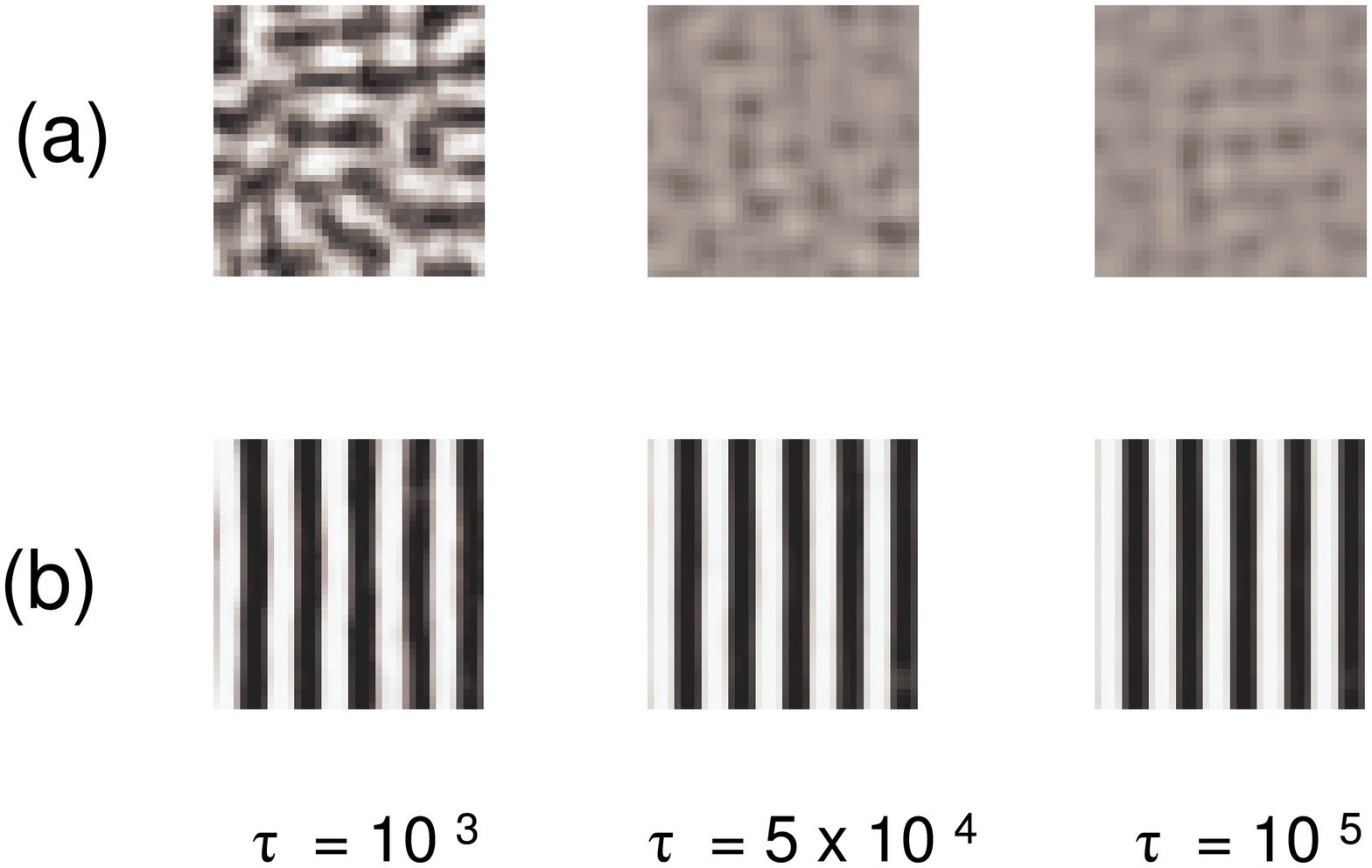}
\caption{Time average of the local perpendicular magnetization $\overline{m_\tau}(\vec{r})$ at different average times $\tau$ (all times are measured in MCS) for  $\delta=3$, $\eta=6.9$ and $L=40$.
Before calculating $\overline{m_\tau}(\vec{r})$ the system was thermalized during $t=10^5$  MCS at each temperature. (a) $T=1.1$ (tetragonal liquid phase); (b) $T=0.9$ (striped phase).}
\label{fig3}
\end{center}
\end{figure}

In figure \ref{fig1} we see that for $\eta > 7$ the system goes through a phase transition
from a phase with perpendicular stripe order, with stripes of width $h=4$ lattice spacings
to a high temperature ``tetragonal phase''. This transition line has been obtained
calculating histograms of the order parameter $O_{hv}$. One such histogram for $\eta=7.5$
and three characteristic temperatures is shown in figure \ref{fig2}. Clear evidence of
a first order transition is given by the behavior of the histogram showing two
metastable phases (stripes and tetragonal) changing stability around the transition
temperature ($ \approx T=1.1$ in this case). This is at variance with results by MacIsaac
{\it et al}.~\cite{MaDeWh1998} who reported a line of second order transitions. The first order
nature of this transitions for large anisotropies is in agreement with recent results
for a corresponding Ising model with dipolar interactions~\cite{CaStTa2004}. Furthermore,
as figure \ref{fig1} shows, this line seems to go asymptotically for large $\eta$, to
a value of the transition temperature $T \approx 1.2$ which is in good agreement with
the phase diagram of the Ising model~\cite{PiCa2007}. This quantitative agreement gives
further credit to the first order nature of this transition line, at least for large $\eta$
where the Ising approximation is justified. More or less direct experimental evidence for
the appearance of stripes magnetic structures was reported already
in an old work by Allenspach and Bischof~\cite{AlBi1992}. More recently the striped nature
of the low temperature phase of high anisotropy, perpendicular Fe/Cu(100) films, together
with the transition to a phase with tetragonal symmetry have been measured and confirmed
by Vaterlaus {\it et al}.~\cite{VaStMaPiPoPe2000}. A theoretical model predicting the existence
of a phase with $90^0$ symmetry induced by the underlying symmetry of the lattice was
put forward by Abanov {\it et al}.~\cite{AbKaPoSa1995}. Abanov {\it et al}. theory works in the Ising
limit where only the perpendicular component of the magnetization is relevant for the
thermodynamic behavior. Their model admits two
possible scenarios for the disordering of the stripes: one similar to the present results
with a first order transition from a stripe phase with positional order decaying
algebraically with distance to a paramagnetic phase with a residual $90^0$ symmetry,
and a second possibility, depending on the values of elastic constants of the theory,
in which an intermediate nematic phase appears between the stripes and paramagnetic phases.
Furthermore, the perpendicular tetragonal phase can evolve continuously to a full
paramagnetic phase or it can finish at a spin reorientation transition. Interestingly,
our phase diagram shows these two behaviors (see figure \ref{fig1}): for an interval
 $6.7 \leq \eta \leq 7$ the Heisenberg system goes from stripes to tetragonal and then
to a planar ferromagnet through a SRT. At still higher temperatures the in-plane
ferromagnet disorders via a continuous transition. This kind of
behavior was already reported in experiments on Fe/Cu(100) ultrathin films by Pappas {\it et al}.~\cite{PaKaHo1990} who found a gap in magnetization between the perpendicular and planar phases
(see figure 1 of Ref.\cite{PaKaHo1990}). Nevertheless, in that early experiments the nature
of the gap was not clear and the authors pointed out two possibilities: a real paramagnetic
gap or the fact that the width of the stripes (not seen in the experiment) could
diminish rapidly in the region of the SRT. One must note that the perpendicular phase
in that series of experiments referred to samples with finite magnetization at low
temperatures, not stripe order. Indeed, further experiments by Allenspach {\it et al}.~\cite{AlBi1992} confirmed the second hypothesis for the same range of thickness of Pappas {\it et al}.. More recently, Won {\it et al}.~\cite{WoWuCh2005} analyzed domain formation and the nature of the SRT
in ultrathin films of Fe/Ni/Cu(001) using high resolution Photoemision Electron Microscopy imaging techniques. They  observed both kind of behaviors, according to the film thickness range, namely, a direct SRT from the striped state and a transition mediated by a paramagnetic gap. In this case, the resolution of the experiment rules out the possibility of domains with a stripe width below the magnetic spatial resolution in the gap region. However, as the authors pointed out, the possibility of a fast-moving striped domain phase cannot be excluded in that experiment. Direct inspection of the typical spin configurations  (see snapshot in figure \ref{fig1}) indicate that in our simulations  the gap corresponds to an out-of-plane tetragonal phase appearing between the perpendicular stripe and planar
ferromagnetic phases. Nevertheless, in experiments only temporal averages can be observed. To emulate the acquisition image process of the experiments, we calculated a time average of the local magnetization (perpendicular component)

\begin{equation}\label{maverage}
    \overline{m_\tau}(\vec{r}) \equiv \frac{1}{\tau} \sum_{t=1}^\tau S^z(\vec{r},t)
\end{equation}

\noindent for different values of the ``acquisition time'' $\tau$, where all the times are measured in MCS. In figure \ref{fig3} we show $\overline{m_\tau}(\vec{r})$ at three different values of $\tau$ in the striped and tetragonal liquid phases. The loss of contrast in the tetragonal liquid phase for relatively short times $\tau$ shows that the characteristic time scales for the fluctuations in this phase are much shorter than in the striped phase.

\begin{figure}
\begin{center}
\hspace{-1cm}\includegraphics[scale=0.32,angle=-90]{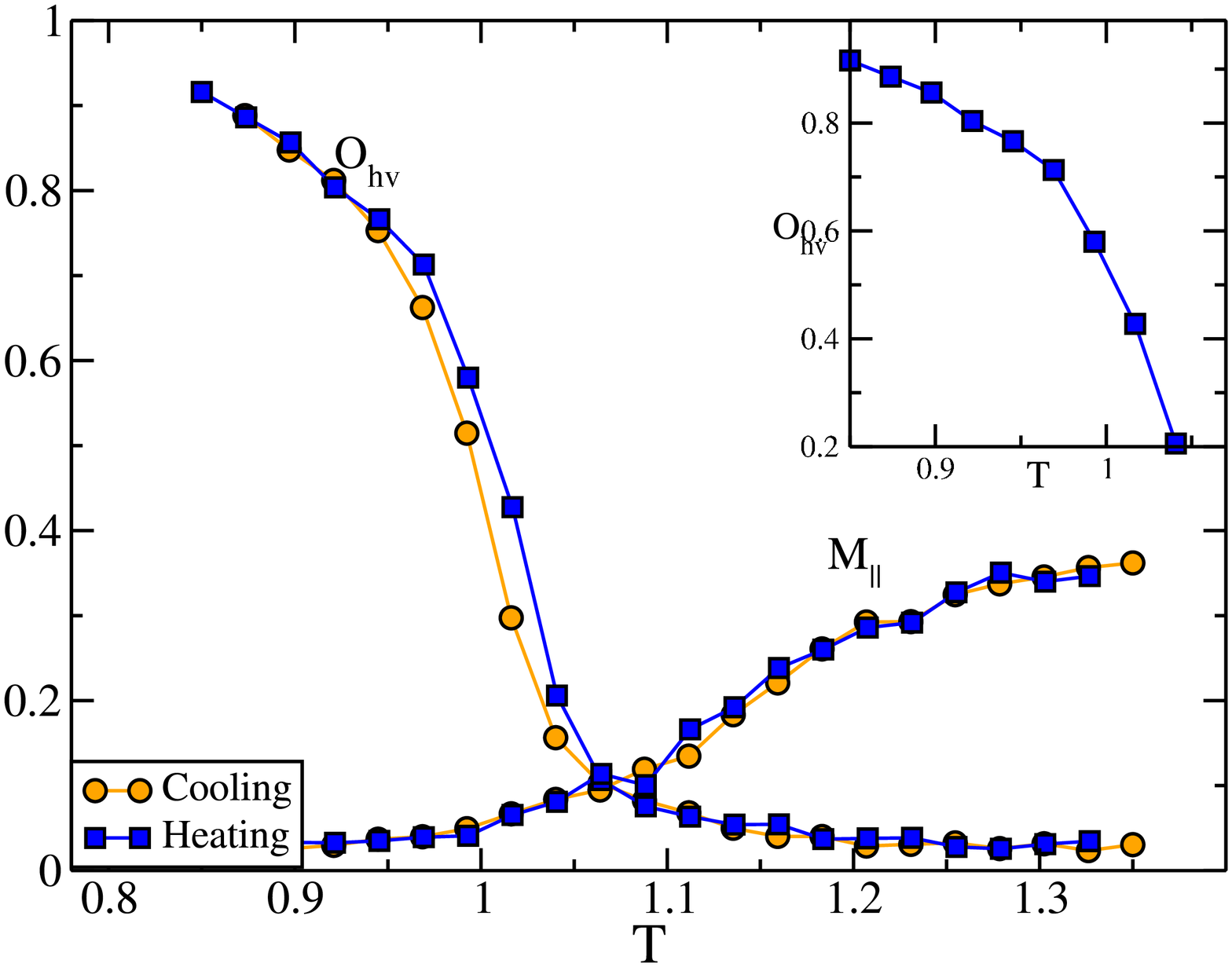}
\caption{(Color online) Order parameter $O_{h\nu}$ and in plane magnetization $M_{||}$ as a function of
temperature for  $\delta=3$ $\eta=6.9$.
The system was first thermalized at $T=1.4$ and cooled with a rate $r= 10^{-7}$ and then heated again with the same rate. The error bars are of the same order of the symbol size. The inset shows a zoom of the heating curve}
\label{fig4}
\end{center}
\end{figure}

In figure \ref{fig4} we show the orientational order parameter $O_{hv}$ and
in-plane magnetization $M^{||}$ for $\eta=6.9$ as a function of
temperature.  Figure \ref{fig4} was obtained by performing cooling and heating
cycles at a very small cooling rate $r=10^{-7}$. Note that the stripe-tetragonal transition
shows weak hysteresis. This indicates that the transition may be weakly first order or
even continuous in this region.
For giving a definite answer one must simulate larger samples and obtain
much better statistics. Nevertheless it is clear from these curves that the sharp first
order transition present for higher values of the anisotropy is much weaker in this region.
Notice also the presence of a small shoulder in  the orientational order parameter (see inset of figure \ref{fig4}). This  effect is more marked in many individual realizations of the stochastic noise, where  an almost  saturated value  of $O_{hv}$  smaller than one can be observed in a narrow range of temperatures below the transition one. The same effect appears for larger  values of $\eta$. This opens the possibility for the second scenario predicted by Abanov {\it et al}.~\cite{AbKaPoSa1995} of an intermediate perpendicular nematic phase,
with long range orientational order but without positional order. Evidence for this
phase comes also from simulations of the Ising dipolar model~\cite{CaMiStTa2006} and
a recent theoretical model for the nematic transition in two
dimensional systems with competing interactions~\cite{BaSt2007,NiSt2007}.

\subsection{Spin Reorientation  Transition}

\begin{figure}
[hb!]
\begin{center}
\hspace{-1cm}\includegraphics[scale=0.35,angle=-90]{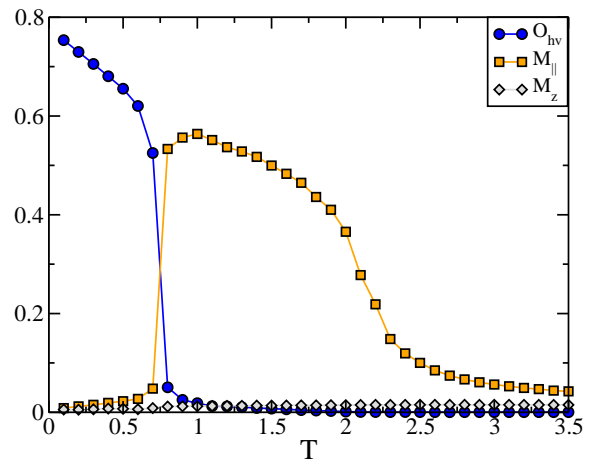}
\caption{(Color online) Order parameter $O_{h\nu}$, in plane and out of plane magnetizations $M_{||}$ and $M_z$ as a function of
temperature for  $\delta=3$ and $\eta=6.5$.
The system was initialized at infinite temperature, thermalized  at $T=6$ and then
it was cooled (equilibrating at each temperature).}
\label{fig5}
\end{center}
\end{figure}

In the region between $\eta=5.8$ and $\eta=7.0$ we observe a sharp SRT
directly from a perpendicular stripe phase to a planar ferromagnetic
one. The behavior of the orientational order parameter and the
in-plane magnetization with temperature is shown in figure \ref{fig5}
for $\eta=6.5$. In this region there is no gap between the
perpendicular and in-plane phases.

\begin{figure}
\begin{center}
\hspace{-1cm}\includegraphics[scale=0.35,angle=-90]{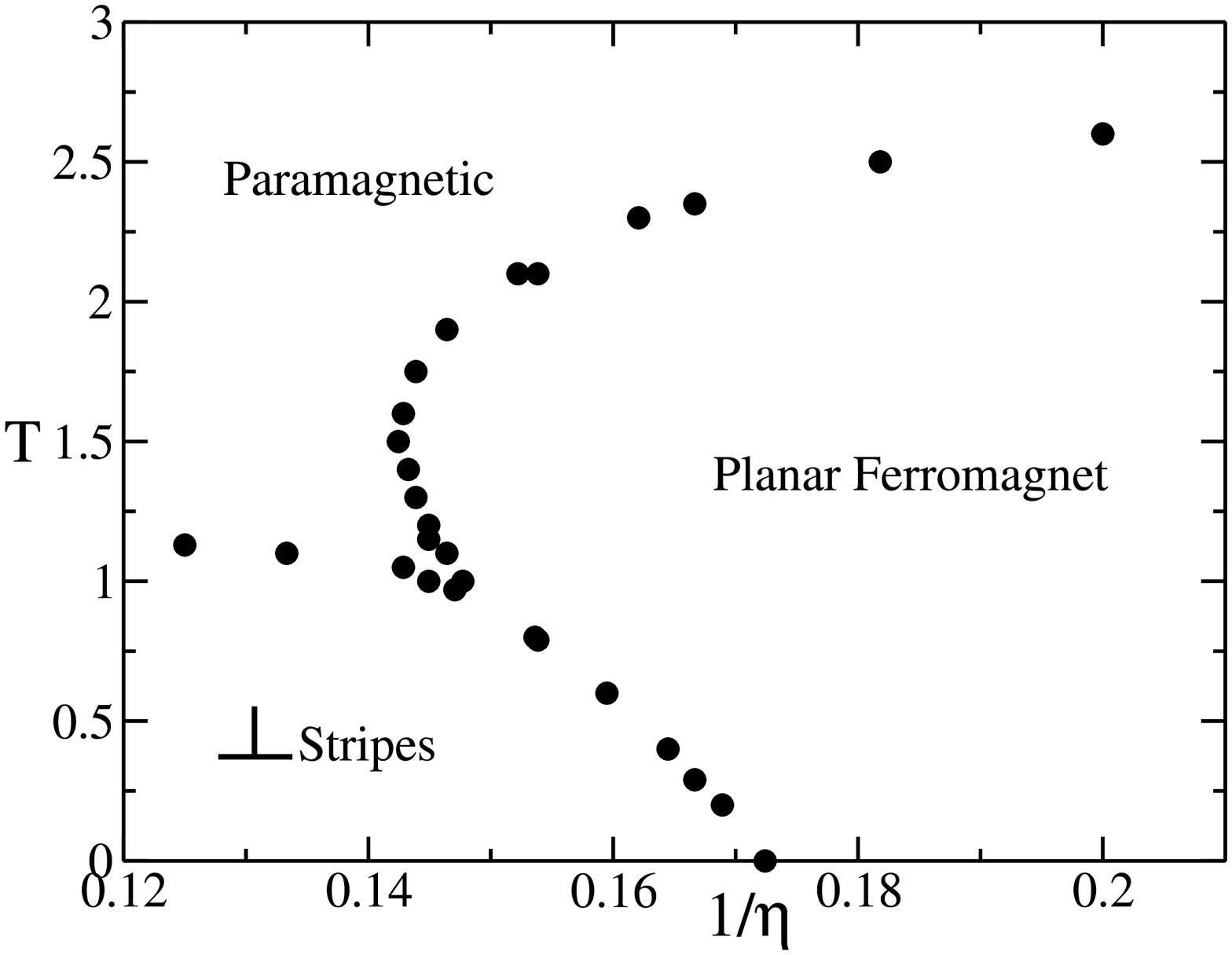}
\caption{Phase diagram temperature versus inverse anisotropy}
\label{fig6}
\end{center}
\end{figure}

The SRT can be accessed both by varying the temperature in a film of
fixed thickness or also by varying the thickness $d$ of the film at fixed
 temperature. In fact, as the thickness grows the in-plane
anisotropy induced by the dipolar interactions is reinforced, while
the perpendicular anisotropy stays nearly constant due to its
essentially surface character. Consequently at some thickness a
SRT can be observed. Then, it is reasonable to consider a
phenomenological model where the thickness acts equivalently to the
inverse anisotropy: $d \propto 1/\eta$.
In fact, Won {\it et al}.~\cite{WoWuCh2005} reported detailed measurements of
magnetic changes as the thickness or temperature of samples of
Fe/Ni/Cu(001) changed. They rationalized the observed
behavior through a phenomenological model and summarized their
findings in a phase diagram ``temperature versus Fe thickness'',
figure 5 of the cited paper. Assuming an approximate equivalence
between thickness and inverse anisotropy, as explained above, we
plotted our simulation data in a ``$T$ versus $1/\eta$'' diagram, as
shown in figure \ref{fig6}.
This figure shows a striking similarity with the right half of figure
5 of Won {\it et al}.
reinforcing the equivalent character of film width and anisotropy in these
systems.

 The
 order of appearance of the perpendicular and planar phases is the
 main difference between our results and a previous phase diagram for
 the same model obtained by MacIsaac {\it et al}.~\cite{MaDeWh1998}. Those
 authors obtained a SRT line in the reverse order, from perpendicular
 at high temperatures to planar at low temperatures. As shown above, the correctness
 of our results is supported by experimental evidence on different
 ultrathin films as well as by theoretical analysis on the effect of
 thermal fluctuations on the SRT at fixed film thickness. Thermal
 fluctuations renormalize the dipolar and anisotropy coupling
 parameters in such a way that the anisotropy $K(T)$ diminishes faster
 than the dipolar coupling constant
 $g(T)$~\cite{PePo1990,PoRePi1993} (in our notation $\eta=K/g$).
Those works predict a linear
 dependence of the transition temperature $T_{SRT}(\eta)$ with
 anisotropy with positive slope, which is roughly in agreement with
 our SRT line from Monte  Carlo simulations.
\begin{figure}
\begin{center}
\hspace{-1cm}\includegraphics[scale=0.32,angle=-90]{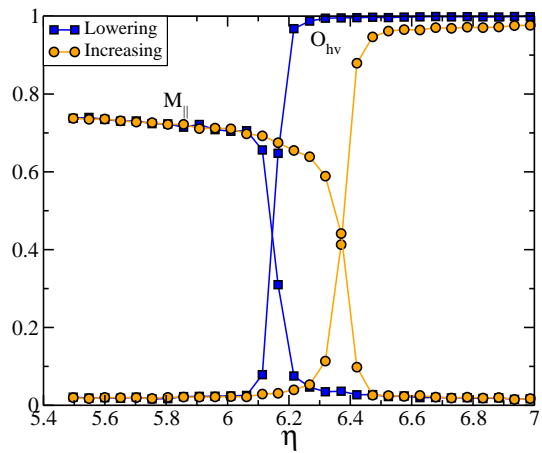}
\caption{(Color online) Order parameter $O_{h\nu}$ and in plane magnetization $M_{||}$ as a function of
$\eta$ for  $\delta=3$ and $T=0.6$.
The system was initialized in stripes of with 4, thermalized  at $\eta=7.5$ and then
$\eta$ was lowered and then increased with a rate of $r=10^{-5}$.}
\label{fig7}
\end{center}
\end{figure}
In figure \ref{fig7} we show cycles of $O_{hv}$ and $M^{||}$ varying
$\eta$ at a fixed temperature of $T=0.6$ in the SRT region. The cycles
show a strong hysteretic behavior. This is further confirmed by means
of energy histograms shown in figure \ref{fig8} which show again the
change in stability between
the perpendicular and planar phases, a signature of the first order
nature of the SRT, as predicted theoretically by several authors~\cite{PePo1990,Po1998}.
\begin{figure}
\begin{center}
\hspace{-1.5cm}\includegraphics[scale=0.3,angle=-90]{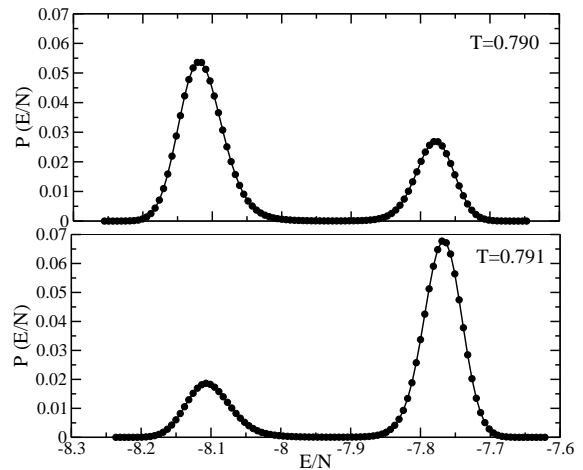}
\caption{Energy per spin histograms for $\delta=3$, $\eta=6.5$ and $T=0.790$ (up)
and $T=0.791$ (down). The histograms were calculated for $30 \times 10^6$ energies measured along a single MC run.}
\label{fig8}
\end{center}
\end{figure}
\subsection{Planar Ferromagnetic-Paramagnetic Transition}
This transition line shows a maximum around $\eta=7$, $T=1.5$ in the
($\eta,T$) plane. One can expect that the behavior of the system across
the transition line will be different in the regions to the right and
to the left of the maximum point. In the far right the system goes
continuously from an in-plane ferromagnet to a paramagnetic phase.
\begin{figure}
\begin{center}
\hspace{-1.5cm}\includegraphics[scale=0.35,angle=-90]{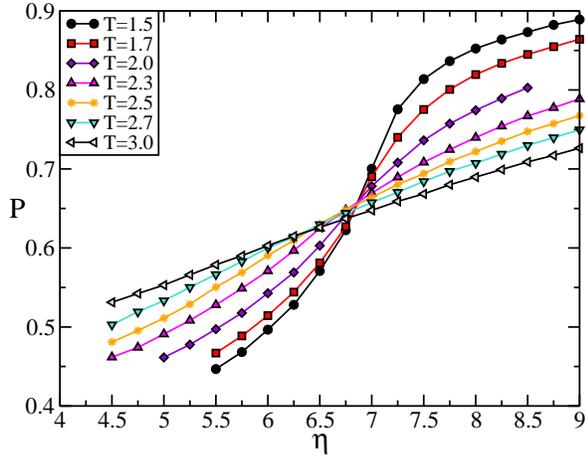}
\caption{(Color on line) Mean absolute magnetization $P$ for different temperatures as function of $\eta$
  across the planar ferro-paramagnetic line ($\delta=3$ and $L=24$).}
\label{fig9}
\end{center}
\end{figure}
We have not been able to characterize completely the nature of this
transition, although our results appear to be consistent with a second order one.

 In figure \ref{fig9} we show that already at small values of
 $\eta$, deep in the planar phase, the spins have a finite
 perpendicular component, which grows continuously with $\eta$ as the
 system goes through the phase transition. At some point around
 $\eta=7$ the curves show an inflexion point upon which the
 perpendicular component tends to saturate. This value of $\eta$
 drifts towards slightly smaller values as the temperature
 increases. We do not have a clear interpretation for the crossing
 points. It would be very
 interesting to analyze the domain walls in the paramagnetic phase and
 how they influence the evolution of the perpendicular magnetization
as the system goes through the phase transition with finite in-plane
magnetization. One may naively expect the perpendicular component of
the local magnetizations $m_i^z$ to vanish in this region, but this
is not the case as figure \ref{fig9} shows.

\begin{figure}
\begin{center}
\hspace{-1.5cm}\includegraphics[scale=0.35,angle=-90]{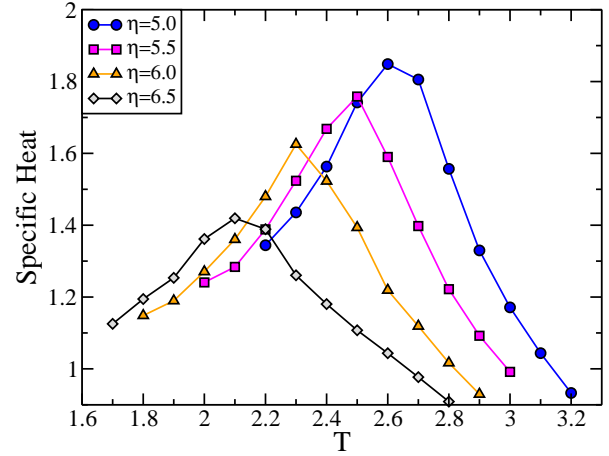}
\caption{(Color on line) Specific heat for $\delta=3$ and different values of $\eta$.}
\label{fig10}
\end{center}
\end{figure}

\begin{figure}
\begin{center}
\hspace{-1.5cm}\includegraphics[scale=0.35,angle=-90]{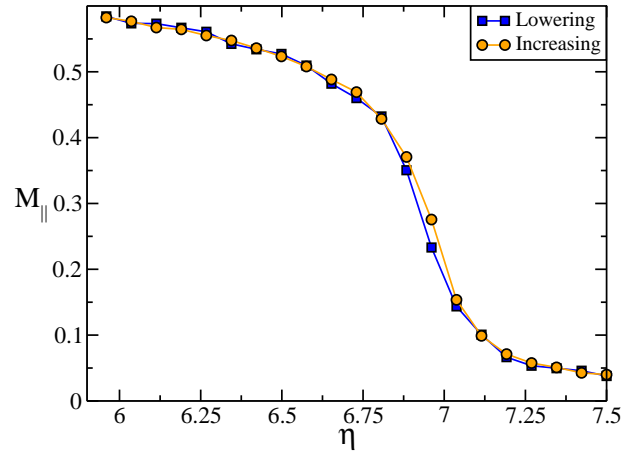}
\caption{(Color on line)  $M_{||}$ as a function of $\eta$ for a fixed Temperatures $T=1.3$.}
\label{fig11}
\end{center}
\end{figure}

In figure \ref{fig10} specific heat curves are shown for different
anisotropies in the same region to the right of the maximum along the
transition line. Note that the peak in the specific heat decreases as
the transition approaches the maximum point from the right, suggesting
a weakening of the second order character of the transition in this
direction.
In figure \ref{fig11} we show an hysteresis cycle in $\eta$ of the
parallel component of the magnetization for a temperature to the left
of the maximum point of the curve. A very weak hysteresis effect is
observed. This behavior is compatible with a continuous transition or
even a weakly first order one. Note that above the transition line in
this region the system enters the tetragonal phase as discussed
above. This phase has a different symmetry as compared with the
paramagnetic high temperature phase. The change from the continuous rotational symmetry of the planar ferromagnetic phase to the discrete rotational symmetry of the tetragonal liquid would be compatible with a discontinuous phase transition in that part of the phase diagram. Also notice that actually  along this line there is also a SRT, because the tetragonal liquid phase is perpendicularly oriented. Since  we already showed the first order nature of the SRT at planar ferromagnetic--striped transition line (where a similar change of symmetry happens), one would expect the planar ferromagnetic--tetragonal liquid line to present the same character. Indeed, a mean field analysis of a multilayered version of the model\cite{MoUs1995} predicts  first order  SRT in the monolayer limit. The previous analysis seem to indicate that a
different nature of the phase transition to the left and right of the
maximum is possible, although  more detailed studies are necessary
in order to elucidate this point.

\section{Pure Dipolar plus Anisotropy film}
\label{delta0}

\begin{figure}
\begin{center}
\includegraphics[scale=0.32]{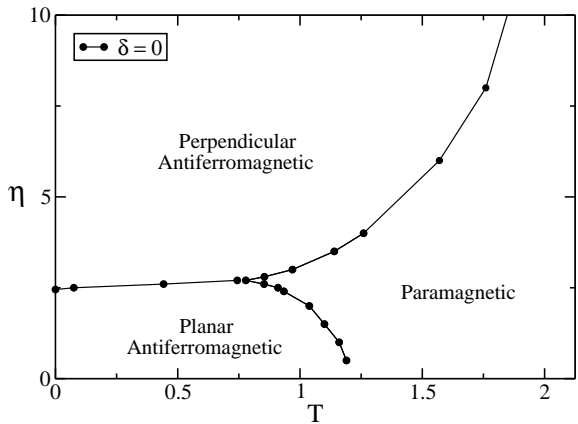}
\caption{Phase diagram for  $\delta=0$ (pure dipolar plus anisotropy)}
\label{fig12}
\end{center}
\end{figure}

In this section we briefly discuss results for the limit where
exchange interactions are absent, $\delta=0$. Experimentally this
limit may be relevant for the behavior of arrays of
magnetic monodomain particles for application in data storage
devices. Usually these arrays can be considered as composed of
noninteracting dots, but as the density of dots grows dipolar effects
may begin to be relevant for the magnetic behavior. Although
relaxation effects of arrays of this type have been studied by several
authors, much less is known on the thermodynamic properties of the
system. In particular, MacIsaac {\it et al}.~\cite{MaWhDePo1996} obtained a phase diagram by
Monte Carlo simulations. Without exchange interactions the relevant
ordered phases in this case are all antiferromagnetic: one out of plane, with sublattice
magnetization and the other one in-plane\cite{MaWhDePo1996,DeMaBoWh1997}. A SRT is also found in this limit, from a planar.
antiferromagnetic phase at small anisotropies to a perpendicular
antiferromagnetic phase at large anisotropies. Similar to what
happened in the $\delta\neq 0$ case, MacIsaac {\it et al}. also found a
reverse order of appearance of the phases through the SRT with temperature
(see figure
1 of Ref.\cite{MaWhDePo1996}). We obtained instead a different behavior,
again similar to the trend of the $\delta\neq 0$ case, from
perpendicular at low temperatures to planar at high temperatures, as
shown in figure \ref{fig12}.

\begin{figure}
\begin{center}
\includegraphics[scale=0.3]{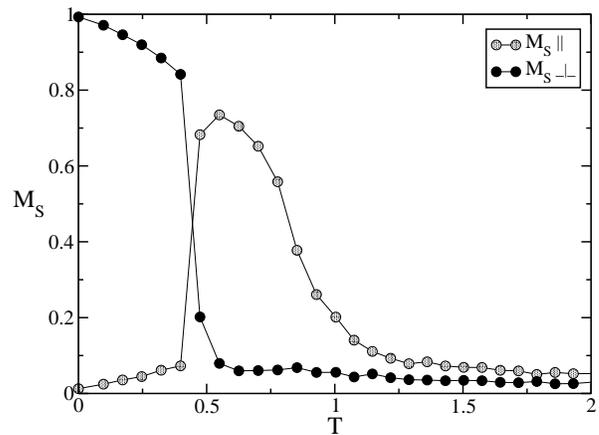}
\caption{Order parameters as function of temperature for $\delta = 0$
  and $\eta=2.6$.
Here the system size is $L=32$.}
\label{fig13}
\end{center}
\end{figure}

Comparing with figure \ref{fig1} we can note that the slope of the
SRT line is very small. Nevertheless there is a finite window where the
transition from out of plane sublattice magnetization to in-plane is
sharp as can be seen in figure \ref{fig13}.
Another important difference between the phase diagrams of figures
\ref{fig1} and \ref{fig12} is the absence of the gap for any fixed
anisotropy in the latter case. This may be related with the different
symmetry of the phases in the pure dipolar case. Now there is no
tetragonal phase and for large anisotropies the system goes directly
from a perpendicular antiferromagnetic phase to a perpendicularly
disordered phase with full rotational symmetry. In this sense the
paramagnetic phase shows the same symmetry along the whole
planar-paramagnetic line at variance with the corresponding line in
the ferromagnetic case.
\section{Conclusions}
In this work we have analyzed the finite temperature phase diagram of
a model for ultrathin ferromagnetic films with exchange, dipolar and
perpendicular
anisotropy interactions for two different values of the exchange
constant
(relative to the dipolar one). Particular emphasis was put in the $\delta=3$ case, where the system presents a striped phase of width $h=4$ at low temperatures and a SRT to a planar ferromagnetic phase as the temperature increases. Although we were not able to simulate systems with larger (and more realistic) values of the exchange constant, the overall qualitative good agreement with many experimental results indicates that the same global behavior should be expected. In particular, the comparison between our phase diagram and the temperature {\it vs.} film thickness of Won {\it et al}. for Fe/Ni/Cu films~\cite{WoWuCh2005} suggests that the film thickness acts as an effective inverse anisotropy. We also reproduced the gap between the striped and the planar phases found by those authors. Moreover, our results indicate that the physical origin of the gap relies in the presence of a fast--moving perpendicularly--oriented labyrinthine (tetragonal liquid) phase.  Evidence of a similar phenomenon (a fast moving striped phase close to the order-disorder transition) in Fe on Cu films has been reported by Portmann {\it et al}.\cite{PoVaPe2006}.

Concerning the thermodynamical nature of the different transitions involved in the phase diagram, we obtained a clear numerical evidence of a first order stripe-planar SRT at low temperatures.  We also found evidence pointing toward a first order nature of the stripe-tetragonal liquid  transition, consistent with previous results in the Ising (i.e., high anisotropy) limit\cite{CaStTa2004,CaMiStTa2006}.

The planar ferromagnet-disordered transition line presents a maximum
in the $(\eta,T)$ space. In the left part of this line, the disordered
state is a tetragonal liquid state, while in the right part we have a
transition to an isotropic paramagnetic state; above the maximum the
system passes continuously (i.e., without any thermodynamical phase
transition) from the tetragonal liquid to the paramagnet. This fact,
together with several other physical arguments, suggests the
possibility of a change in the order of the transition around the maximum of the line, being weakly first order in the left part of the line and second order in the right part. If confirmed, this would imply the existence of a tricritical point around the maximum and a triple point where the three phases (stripe-planar-tetragonal) coexist. However, strong finite size effects did not allow us to give a definite answer concerning this point and further studies should be needed.

In the case of $\delta=0$ we showed the existence of a SRT from a perpendicular antiferromagnetic phase at low temperature to an in-plane antiferromagnetic phase at higher temperatures, at variance with previous reported results. The present results suggest that a SRT from a low temperature out-of plane to an in-plane phase at higher temperatures for low values of $\eta$ is present for any value of $\delta$.

This work was partially supported by grants from
CONICET (Argentina), SeCyT, Universidad Nacional de C\'ordoba
(Argentina), CNPq (Brazil), FONCyT grant PICT-2005 33305 (Argentina) and ICTP
grant NET-61 (Italy).


\begin{thebibliography}{10}

\bibitem{Portmann2006}
O. Portmann, {\em Micromagnetism in the Ultrathin Limit} (Logos Verlag,
  ADDRESS, 2006).

\bibitem{PaKaHo1990}
D.~P. Pappas, K.~P. Kamper, and H. Hopster, Phys. Rev. Lett. {\bf 64},  3179
  (1990).

\bibitem{AlBi1992}
R. Allenspach and A. Bischof, Phys. Rev. Lett. {\bf 69},  3385  (1992).

\bibitem{YaGy1988}
Y. Yafet and E.~M. Gyorgy, Phys. Rev. B {\bf 38},  9145  (1988).

\bibitem{WoWuCh2005}
C. Won {\it et~al.}, Phys. Rev. B {\bf 71},  224429  (2005).

\bibitem{VaStMaPiPoPe2000}
A. Vaterlaus {\it et~al.}, Phys. Rev. Lett. {\bf 84},  2247  (2000).

\bibitem{MaWhRoDe1995}
A.~B. MacIsaac, J.~P. Whitehead, M.~C. Robinson, and K. De'Bell, Phys. Rev. B
  {\bf 51},  16033  (1995).

\bibitem{CaStTa2004}
S.~A. Cannas, D.~A. Stariolo, and F.~A. Tamarit, Phys. Rev. B {\bf 69},  092409
   (2004).

\bibitem{CaMiStTa2006}
S.~A. Cannas, M. F. Michelon, D.~A. Stariolo, and F.~A. Tamarit, Phys. Rev. B {\bf
  73},  184425  (2006).

\bibitem{RaReTa2006}
E. Rastelli, S. Regina, and A. Tassi, Phys. Rev. B {\bf 73},  144418  (2006).

\bibitem{MaDeWh1998}
A.~B. MacIsaac, K. De'Bell, and J.~P. Whitehead, Phys. Rev. Lett. {\bf 80},
  616  (1998).

\bibitem{NiSt2007}
L. Nicolao and D.~A. Stariolo, Physical Review B (Condensed Matter and
  Materials Physics) {\bf 76},  054453  (2007).

\bibitem{BaSt2007}
D.~G. Barci and D.~A. Stariolo, Physical Review Letters {\bf 98},  200604
  (2007).

\bibitem{AbKaPoSa1995}
A. Abanov, V. Kalatsky, V.~L. Pokrovsky, and W.~M. Saslow, Phys. Rev. B {\bf
  51},  1023  (1995).

\bibitem{BoMaWhDe1995}
I. Booth, A.~B. MacIsaac, J.~P. Whitehead, and K. De'Bell, Phys. Rev. Lett.
  {\bf 75},  950  (1995).

\bibitem{MaWhDePo1996}
A.~B. MacIsaac, J.~P. Whitehead, K. De'Bell, and P.~H. Poole, Phys. Rev. Lett.
  {\bf 77},  739  (1996).

\bibitem{DeMaBoWh1997}
K. De'Bell, A.~B. MacIsaac, I.~N. Booth, and J.~P. Whitehead, Phys. Rev. B {\bf
  55},  15108  (1997).

\bibitem{PiCa2007}
S.~A. Pighin and S.~A. Cannas, Physical Review B (Condensed Matter and
  Materials Physics) {\bf 75},  224433  (2007).

\bibitem{PePo1990}
D. Pescia and V.~L. Pokrovsky, Phys. Rev. Lett. {\bf 65},  2599  (1990).

\bibitem{PoRePi1993}
P. Politi, A. Rettori, and M.~G. Pini, Phys. Rev. Lett. {\bf 70},  1183
  (1993).

\bibitem{Po1998}
P. Politi, Comments Cond. Matter Phys. {\bf 18},  191  (1998).

\bibitem{MoUs1995}
A. Moschel and K.~D. Usadel, Phys. Rev. B {\bf 51},  16111  (1995).

\bibitem{PoVaPe2006}
O. Portmann, A. Vaterlaus, and D. Pescia, Phys. Rev. Lett. {\bf 96},  047212
  (2006).

\end{thebibliography}
\end{document}